\documentclass[aps,twocolumn,psfig,graphicx,amsfonts,showpacs,tighten,
eqsecnum,floats,amssymb,amsmath]{revtex4}
\usepackage{graphicx}

\makeatother

\def\beq{\begin{eqnarray}}
\def\eeq{\end{eqnarray}}
\def\nn{\nonumber\\}

\def\ub#1{\underleftarrow{#1}}

\def\mbf#1{\mbox{\boldmath${#1}$}}
\def\del{\partial}
\def\grad{\nabla}
\def\lie{\pounds}
\def\half{{\textstyle{1\over2}}}

\begin{document}
\title{Generic weak isolated horizons}

\author{Ayan Chatterjee}\email{ayan.chatterjee@saha.ac.in}
\author{Amit Ghosh}\email{amit.ghosh@saha.ac.in}

\affiliation{Theory Division, Saha Institute of
Nuclear Physics, Kolkata 700064, India}

%%%%%%%%%%%%%%%%%%%%%%%%%%%%%%%%%%%%%%%%%%%%%%%%%%%%%%%%%%%%%%%%%%%%%%
\begin{abstract}

Weak isolated horizon boundary conditions have been relaxed supposedly to their
weakest form such that both zeroth and the first law of black hole mechanics still
emerge, thus making the formulation more amenable for applications in both analytic
and numerical Relativity. As an additional gain it explicitly brings the non-extremal 
and extremal black holes at the same footing.

\end{abstract}

\pacs{04070B, 0420}

\maketitle

%%%%%%%%%%%%%%%%%%%%%%%%%%%%%%%%%%%%%%%%%%%%%%%%%%%%%%%%%%%%%%%%%%%%%%%%

Isolated horizons \cite{ash1}, much like Killing horizons \cite{wald1}, were
introduced in order to deal with the more practical problems involving black holes.
Both formulations are {\em local}, in contrast with the global nature of event
horizons \cite{bch}. However, isolated horizons, unlike Killing horizons, do 
not require a Killing vector field in their neighbourhood. Thus,
isolated horizons are characterized by a weaker set of boundary conditions that 
give rise to both the zeroth and the first laws of black hole mechanics. 
Nevertheless, the isolated horizon boundary conditions, as developed in a 
series of papers \cite{ash1,ash2,ash3,afk,ash4,ash5}, consider
a restricted equivalence class of null normal vector fields $[\,c\ell^a]$, where
$\ell^a_1,\,\ell^a_2$ belong to the class if and only if $\ell^a_1=c\ell^a_2$ for
some positive definite constant $c$. Even though it has been emphasized that the
most natural equivalence class of null normals at the horizon is $[\,\xi\ell^a]$,
where $\xi$ is an arbitrary positive function on the horizon, one restricts oneself,
somewhat artificially perhaps, to the constant equivalence class. The purpose of
this letter is to generalize the framework such that now the equivalence class of
null normals is $[\,\xi\ell\,]$, where $\xi$ is a given class of functions to be 
specified below, and to show how from this generalized setup both the zeroth and the
first laws of black hole mechanics can be derived. This clearly makes isolated
horizons applicable to a wider varieties of problems, both from analytical and
numerical viewpoints. Another unexpected gain is that this generalized class of null
normals include both non-extremal and extremal global solutions explicitly, namely
there exists a choice of $\xi$ for which the surface gravity associated with the
vector field $\xi\ell^a$ vanishes, even though it is nonzero for other null normals
in the class. In contrast, Killing horizons make essential use of the bifurcation two
spheres in proving the zeroth and the first law \cite{wald2}, thereby distinguish extremal 
and non-extremal solutions in an explicit way.   
%%%%%%%%%%%%%%%%%%%%

Let $\cal M$ be a four-manifold equipped with a metric $g_{ab}$ of signature
$(-,+,+,+)$. Our notations and conventions closely follow that of \cite{afk}.
${\Delta}$ is a null hypersurface in $\cal M$ of which $\ell^a$ is a future directed null
normal. However, if $\ell^a$ is a null normal, so is $\xi\ell^a$, where $\xi$ is any
positive function on $\Delta$. Thus, $\Delta$ naturally admits an equivalence class
of null normals $[\,\xi\ell\,]$. We denote by $q_{ab}\triangleq g_{\ub{ab}}$ the
degenerate intrinsic metric on $\Delta$ induced by $g_{ab}$ (indices that are not
explicitly intrinsic on $\Delta$ will be pulled back and $\triangleq$ means that
the equality holds {\em only on} $\Delta$). A tensor $q^{ab}$ will be an {\em inverse} of
$q_{ab}$ if it satisfies $q^{ab}q_{ac}q_{bd}\triangleq q_{cd}$. The expansion
$\theta_{(\ell\,)}$ of the null normal $\ell$ is then defined by
$\theta_{(\ell\,)}=q^{ab}\nabla_a\ell_b$,
where $\nabla_a$ is the covariant derivative compatible with $g_{ab}$.

\vskip 0.5cm
\noindent {\em Boundary conditions and zeroth law}\,: $\Delta$ is a non-expanding
horizon (NEH) in $({\cal M},g_{ab})$ if the following conditions are satisfied: {\tt
1)} Topologically $\Delta\equiv\mathbb S^2\times\mathbb R$, {\tt 2)}
$\theta_{(\xi\ell\,)}\triangleq 0$ for any null normal $\xi\ell$ in the class
$[\,\xi\ell\,]$ (actually, this is just {\em one} condition, since
$\theta_{(\xi\ell\,)}\triangleq\xi\theta_{(\ell\,)}$, it suffices that
$\theta_{(\ell\,)}\triangleq 0$, {\tt 3)} all equations of motion including matter
hold on $\Delta$, in particular the matter energy-momentum tensor $T_{ab}$ is such
that $-T^a{}_b\,\xi\ell^b$ is future directed and causal on $\Delta$. The important
thing to note here is that all boundary conditions are intrinsic to $\Delta$.

We shall always work with the null tetrad basis $(\ell, n, m, \bar{m})$ such that
$1\!=\!-n\cdot\ell=\!m\cdot\bar m$ and all other scalar products vanish. Since any null
normal $\xi\ell$ is expansion-free, twist-free and geodetic, {\em i.e.}
$\xi\ell^a\grad_a(\xi\ell^b)\triangleq\kappa_{(\xi\ell\,)}\xi\ell^b$, where
$\kappa_{(\xi\ell\,)}$ is the acceleration of $\xi\ell^a$, the Raychaudhuri equation becomes
\begin{equation}\label{eqray}
0\triangleq\lie_{\xi\ell}\,\theta_{(\xi\ell\,)}\triangleq-|\sigma_{(\xi\ell\,)}|^2
-\xi^2R_{ab}\ell^a\ell^b
\end{equation}
where, $\sigma_{(\xi\ell\,)}=m^am^b\nabla_a(\xi\ell_b)$ is the shear of $\xi\ell^a$.
Since by energy condition $R_{ab}\ell^a\ell^b$ is positive both terms on the right
hand side vanish independently on $\Delta$. Therefore, every null normal $\xi\ell^a$ in the
equivalence class is also shear-free.

From these results it follows that a class of one-forms $\omega_a^{(\xi\ell\,)}$ exists on
$\Delta$ such that
\begin{equation}\label{defomega}
\grad_{\ub a}(\xi\ell^b)\triangleq\omega_a^{(\xi\ell\,)}\xi\ell^b
\end{equation}
The one-form varies over the class $[\,\xi\ell\,]$ as
${\omega}^{(\xi\ell\,)}_a=\omega^{(\ell\,)}_a+\grad_{\ub a}\ln\xi$.
It also follows that every null normal $\xi\ell^a$ in the equivalence class
is Killing on $\Delta$, namely $\lie_{(\xi\ell\,)}\,q_{ab}\triangleq 0$.
It is useful to calculate the curvature of $\omega^{(\xi\ell\,)}$ as well.
For the entire class of null-normals we find
\begin{equation} \label{curvomega}
d\omega^{(\xi\ell\,)}\triangleq 2({\rm Im}\Psi_2)\,\mbf{\epsilon}\;,
\end{equation}
where ${\rm Im}\Psi_2=C_{abcd}\ell^am^b\bar m^cn^d$ is a complex scalar constructed
from the Weyl-tensor $C_{abcd}$ and $\mbf{\epsilon}=im\wedge\bar m$.
%%%%%%%%%%%%%%%%%%%%%%%%%%%%%%%%%%%%%%%%%%%%%%%%%%%%%%%%%%%%%%%%%%%

%\subsection{Weak Isolated Horizon}
The acceleration $\kappa_{(\xi\ell\,)}$ varies over the equivalence class
$[\,\xi\ell\,]$ as follows: $\kappa_{(\xi\ell\,)}=\xi\kappa_{(\ell\,)}+\lie_\ell\,
\xi$, hence is not a constants on $\Delta$. In order to obtain the zeroth law we need to
restrict the NEH structure further. The restriction is called the {\em weak isolated
horizon} (WIH) which is a NEH equipped with an equivalence class of null normals
$[\,\xi\ell\,]$ obeying
\beq \label{wih}
\lie_{(\xi\ell\,)}\,\omega^{(\xi\ell\,)}\triangleq 0\;.
\eeq
Two comments are in order at this point. First of all, as in the case of Killing horizons, we will
interpret the acceleration $\kappa_{(\xi\ell\,)}$ as the surface gravity associated
with $\xi\ell^a$. However, a global Killing field being absent, the value of the
surface gravity cannot be uniquely fixed. In isolated horizon formulation it is
natural to maintain this freedom. Secondly, the boundary condition (\ref{wih}) defining a WIH
is not a {\em single} condition representing the entire equivalence class,
since $\lie_{(\xi\ell\,)}\,\omega^{(\xi\ell\,)}\triangleq d(\xi\ell\cdot
\omega^{(\xi\ell\,)})\triangleq d\kappa_{(\xi\ell\,)}$ and $\kappa_{(\xi\ell\,)}$ is
not constant for the entire class. To eliminate this pathology of having an infinite
number of boundary conditions let us restrict our choice of the $\xi$-functions as
follows: given say, $\kappa_{(\ell\,)}\triangleq$ constant on $\Delta$, all
$\kappa_{(\xi\ell\,)}$ are also constants on $\Delta$ if $\xi$ to satisfies the
differential equation
\beq \kappa_{(\xi\ell\,)}\triangleq{\rm
constant}\triangleq\xi\kappa_{(\ell\,)}+\lie_\ell\,\xi\;.\eeq
Choosing a function $v$ such that $\lie_\ell\,v=1$ one can solve the above equation
for $\xi$ obtaining $\xi=\eta\,e^{-v\kappa_{(\ell\,)}}+\kappa_{(\xi\ell\,)}/\kappa_{(\ell\,)}$,
where $\eta$ is a nonzero function satisfying $\lie_\ell\,\eta=0$. In the rest of the
letter, we shall restrict ourselves to $\eta\triangleq$ constant. With this
restriction the boundary condition (\ref{wih}) becomes the {\em single}
representative of the entire equivalence class. This class admits non-extremal as
well as extremal global solutions, since $\xi$ is positive even when
$\kappa_{(\xi\ell\,)}\triangleq 0$.

As already noted, a direct consequence of the WIH boundary condition (\ref{wih}) is
the {\em zeroth law}\,: $d\kappa_{(\xi\ell\,)}\triangleq 0$. Therefore, the surface
gravity corresponding to each $\xi\ell$ in $[\,\xi\ell\,]$ is constant on $\Delta$,
provided $\xi$ is restricted as above.

\vskip 0.5 cm
\noindent{\em First Law}\,: We consider the first order Palatini Lagrangian on $\cal
M$ which has an internal boundary $\Delta$. It is sufficient to consider $\cal M$ to
be bounded by two partial Cauchy surfaces $M_\pm$, both oriented and cuts $\Delta$
at the two-spheres $S_\pm$ (see Figure 1).
\begin{figure}[h] \label{f1}
  \begin{center}
  \includegraphics[height=3.5cm]{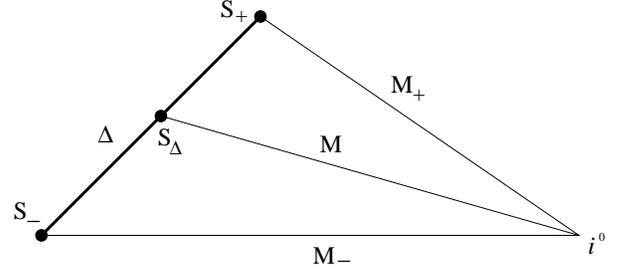}
  \caption{$M_\pm$ are two partial Cauchy
  surfaces enclosing a  region of space-time and 
  intersecting $\Delta$ in the $2$-spheres
  $S_\pm$ respectively and extend to spatial infinity $i^o$. }
  \end{center}
\end{figure}
The basic fields are the co-tetrad $e_a^I$ and the connection one-form $A_{aI}{}^J$.
The lower case Latin letters are spacetime indices while the upper case letters
refer to the internal four dimensional Minkowski spacetime having a fixed metric
$\eta_{I\!J}$ of signature $(-+++)$. The metric on $\cal M$ is given by
$g_{ab}=e_a^Ie_b^J\eta_{I\!J}$. Next we define a covariant derivative operator that
acts only on internal indices
\beq D_av_I=\partial_av_I + {A_{aI}}^J v_J \;, \eeq
where, $\partial$ is a fiducial derivative operator chosen to be torsion free and
compatible with the flat metric. We choose the Lagrangian four-form as
\begin{eqnarray}\label{lagrangian}
L=\frac{1}{16\pi G}\left[-\Sigma^{I\!J}\wedge F_{I\!J}+d(\Sigma^{I\!J}\wedge
A_{I\!J})\right]\;,
\end{eqnarray}
where $\Sigma^{I\!J}=\half\,\epsilon^{I\!J}{}_{KL}e^K\wedge e^L$ and $F=dA+A\wedge
A$.

Next consider the variation of (\ref{lagrangian}). The allowed variations of the
smooth fields $(e^I_a,A)$ on ${\cal M}$ is such that it satisfies the 
standard conditions at infinity to ensure asymptotic flatness and
on $\Delta$ must satisfy the following
conditions: {\tt 1)} each spacetime admits a null normal belonging to the
equivalence class $[\,\xi\ell\,]$, {\tt 2)} $(\Delta,[\,\xi\ell\,])$ is a WIH.

The standard variation gives on-shell $\delta L=d\Theta(\delta)$ where $16\pi G
\Theta(\delta)=\delta\Sigma^{I\!J}\wedge A_{I\!J}$. Following standard procedures
\cite{abr,wald2}, one then constructs the symplectic structure $\Omega$ on the space of
solutions. First, we construct the symplectic current $J(\delta_1,\delta_2)=
\delta_1\Theta(\delta_2)-\delta_2\Theta(\delta_1)$, which is closed on-shell.
Integrating $dJ$ over the spacetime under consideration we find
\beq \int_{M_+\cup M_-\cup\Delta\cup i^0}J=0\;.\eeq
In order to evaluate $J|_\Delta$ we need expressions for the fields on $\Delta$.
Given the null vectors $(\ell,n,m,\bar m)$ the tetrad $e_a^I$ gives the internal
null vectors $(\ell^I,n^I,m^I,\bar m^I)$. In terms of these vectors and with the
help of the fact that the pull-back of the one-form $\ell_{\ub a}\triangleq 0$, the
pull-back of the two-form $\ub\Sigma^{I\!J}$ can be expressed as
\beq\label{sigdelta} &&
\Sigma^{I\!J}_{\ub{ab}}\triangleq 2\ell^{[I}n^{J]}\,\mbf{\epsilon}_{ab}+2i(n_{[\ub
a}m_{\ub b]}\ell^{[I}\bar{m}^{J]}\nonumber\\
&&\qquad\qquad-n_{[\ub a}\bar m_{\ub b]}\ell^{[I}m^{J]}).
\eeq
Using $\del_a\ell_I=0$ and the fact the $\grad$ is compatible with the tetrad
$e_a^I$ we can restrict the connection one-form $A_{I\!J}$ on $\Delta$
\begin{equation}\label{conndelta}
A_{\ub aI\!J}\triangleq-2\ell_{[I}n_{J]}\omega^{(\ell\,)}_a+\alpha_{\ub aI\!J},
\end{equation}
where the one-form $\alpha_{I\!J}$ satisfies the condition $\alpha_{I\!J}\ell^J=0$.
This one-form decouples from rest of the analysis. Formula (\ref{conndelta}) holds
for the entire equivalence class $[\,\xi\ell\,]$ in the following precise sense: We
have picked a fiducial null-normal $\ell$ from the class $[\,\xi\ell\,]$ as our
basis such that the internal null vector $\ell_I$ is annihilated by the derivative
$\del$ and thus arrived at (\ref{conndelta}). But we could have chosen any other
$\xi\ell$ as forming the basis $(\xi\ell,n/\xi,m,\bar m)$. Carrying out the same
analysis as before we would obtain the same decomposition (\ref{conndelta}).

In order to see that $J|_\Delta$ is an exact form we define a potential field
$\psi_{(\xi\ell\,)}$ of the surface gravity $\kappa_{(\xi\ell\,)}$ as\,:
$\lie_{(\xi\ell\,)}\psi_{(\xi\ell\,)}\triangleq\kappa_{(\xi\ell\,)}$. Clearly
$\psi_{(\xi\ell\,)}$ suffers from an additive ambiguity which can be fixed by
requiring that $\psi_{(\xi\ell\,)}|_{S_-}\!\!\triangleq 0$. Using these definitions
and the expressions (\ref{sigdelta}) and (\ref{conndelta}) we find
\beq J(\delta_1,\delta_2)|_\Delta&\triangleq&\frac{1}{8\pi G}\,d\,\big[
\delta_1\psi_{(\ell\,)}\delta_2\mbf{\epsilon}-\delta_2\psi_{(\ell\,)}\delta_1
\mbf{\epsilon}\big]\nn &\triangleq:& dj(\delta_1,\delta_2)\;.\eeq
Taking the orientation of the basis into account, the difference of the integrals
$\int_MJ-\oint_{S_\Delta}j$ is independent of the choice of the Cauchy surfaces $M$
($S_\Delta$ is the two-sphere at which $M$ cuts $\Delta$). Hence, one obtains the
symplectic structure (see Figure 1)
\begin{eqnarray}\label{symplectic} &&
\Omega(\delta_1,\delta_2)=\frac{1}{16\pi G}\int_M\big[\delta_2\Sigma^{I\!J}\wedge
\delta_1A_{I\!J}-\delta_1\Sigma^{I\!J}\wedge\nn &&\!\delta_2A_{I\!J}\big]
-\frac{1}{8\pi G}\oint_{S_\Delta}\big[
\delta_2\mbf{\epsilon}\,\delta_1\psi_{(\ell\,)}-\delta_1
\mbf{\epsilon}\,\delta_2\psi_{(\ell\,)}\big]\;.
\end{eqnarray}
Given the symplectic structure the question now is whether the flow generated by the
vector field $t^a=\xi\ell^a$ is Hamiltonian? To investigate this, we should calculate the
one-form $X_t$ where $X_t(\delta)=\Omega(\delta,\delta_t)$ and $\delta_t$ is to be
interpreted as the Lie-flow $\lie_t$ when acting on geometric fields. Then the
necessary and sufficient condition for the existence of a Hamiltonian $H_t$ is that
the one-form $X_t$ be exact $X_t={\bf d}H_t$, where $\bf d$ is the exterior
differential in the space of solutions. This makes $X_t(\delta)=\delta H_t$.

For the symplectic structure at hand, $X_t(\delta)$ receives contributions both from
the bulk and the surface symplectic structures. The bulk term, thanks to the
equations of motions satisfied by fields $(e_I,A)$ and the linearized equations of
motions satisfied by the variations $(\delta e_I,\delta A)$, contributes to the
two-sphere boundaries of $M$, which are $S_\Delta$ and $S_\infty$. For the bulk term
we get
\begin{eqnarray}\label{oneform}
    X_t(\delta)|_M=
    -\frac{1}{8\pi G}\,\xi\kappa_{(\ell\,)}\delta {\mathcal A}_\Delta+\delta E_t
\end{eqnarray}
where, ${\mathcal A}_\Delta=\int_{S_\Delta}\!\!\mbf\epsilon$ is the area of $S_\Delta$ and $E_t$
is the ADM-energy arising from the integral at $S_\infty$ where it is understood
that the asymptotic time translation is generated by $t^a$.

Care should be taken while calculating $X_t(\delta)$ from the surface symplectic
structure, since one cannot identify $\delta_t\psi_{(\xi\ell\,)}=\lie_t
\psi_{(\xi\ell\,)}$. The reason is $\psi_{(\xi\ell\,)}$ being the potential of
$\kappa_{(\xi\ell\,)}$, its variations are completely determined by those of
$\kappa_{(\xi\ell\,)}$.  Since $\delta_t\kappa_{(\xi\ell\,)}=
\lie_t\kappa_{(\xi\ell\,)}\triangleq 0$ by zeroth law and
$\delta_t\psi_{(\xi\ell\,)}|_{S_-}\triangleq 0$ it turns out that $\delta_t
\psi_{(\xi\ell\,)}\triangleq 0$ everywhere. But
$\psi_{(\xi\ell\,)}=\psi_{(\ell\,)}+\ln \xi$. Thus,
$\delta_t\psi_{(\ell\,)}+\delta_t\ln\xi\triangleq 0$. From $\delta_t(\xi
\ell^a)=\lie_t(\xi\ell^a)=0$ we find that $\delta_t\psi_{(\ell\,)}\triangleq
-\lie_\ell\,\xi$. Now we are in a position to calculate $X_t(\delta)$ arising from
the surface symplectic structure
\begin{eqnarray}\label{oneformgrav}
   X_t(\delta)|_{S_\Delta}=-\frac{1}{8\pi G}\lie_{\ell}\xi~\delta {\mathcal A}_{\Delta}
\end{eqnarray}
Combining (\ref{oneform}) and \ref{oneformgrav} we get
\beq X_t(\delta)\triangleq -\frac{1}{8\pi G}\,\kappa_{(\xi\ell\,)}\delta {\mathcal A}_{\Delta}+
\delta E_t\;.\eeq
This gives the first-law $\delta E_\Delta^t=\kappa_{(\xi\ell\,)}\delta{\mathcal
A}_\Delta/8\pi G$ once we identify $X_t={\bf d}H_t$ and define $E_t-H_t=E_\Delta^t$, where
$E_\Delta^t$ is the horizon-mass. That $X_t$ is a closed form forces one to regard
$\kappa_{(\xi\ell\,)}$ as a function of the area ${\mathcal A}_\Delta$ alone.

\vskip 0.5cm
\noindent\textit{Electromagnetic Field}\,: NEH boundary conditions restricts
the matter energy-momentum tensor by $T_{ab}\ell^a\ell^b\triangleq 0$. For the
electromagnetic fields $\bf A$ this implies
\begin{equation}\label{emtll}
0\triangleq T_{ab}\ell^a\ell^b\triangleq|\,\ell^am^b{\bf F}_{ab}|^2\;,
\end{equation}
where ${\bf F}=d{\bf A}$ is the electromagnetic field strength. It follows from
(\ref{emtll}) that $\ell^a{\bf F}_{a\ub b}\triangleq 0$ and $\ell^a\!*{\bf
F}_{a\ub b}\triangleq 0$. These two restrictions tell us there is no flux of
electromagnetic radiation {\em across} the horizon. Radiation may flow along the
horizon. Since $\Delta$ is an inner boundary and the normal to $S_\Delta$ is inward
pointing, the electric charge of the horizon is defined by (it is assumed that all
magnetic charges are zero)
\begin{equation}\label{charge}
Q_{\Delta}\triangleq-\frac{1}{4\pi}\oint_{S_\Delta}\!\!\! *{\bf F}\;.
\end{equation}
Since $\lie_{(\xi\ell\,)}\ub{*{\bf F}}\triangleq 0$ the charge $Q_\Delta$ is independent
of the choice of $S_\Delta$.

\vskip 0.5cm
\noindent{\em Zeroth Law}\,: To establish the zeroth law for the electromagnetic case,
we need to define an electric potential $\Phi$ at the horizon. For this the
electromagnetic potential ${\bf A}$ is gauge fixed at the WIH such that
\begin{equation}\label{lieema}
\lie_{(\xi\ell\,)}{\bf A}_{\ub a}\triangleq\grad_{\ub a}\lie_{(\xi
\ell\,)}\chi_{(\xi\ell\,)}\;.
\end{equation}
where $\chi_{(\xi\ell\,)}$ is a fixed function of $v$. Given such an electromagnetic
potential ${\bf A}$ we can now define the scalar potential $\Phi_{(\xi\ell\,)}$ at
the horizon as
\beq \Phi_{(\xi\ell\,)}\triangleq -\xi\ell\cdot{\bf
A}+\lie_{(\xi\ell\,)}\chi_{(\xi\ell\,)}\;. \eeq
From (\ref{lieema}) it follows immediately that $d\Phi_{(\xi\ell\,)}\triangleq 0$,
hence $\Phi_{(\xi\ell\,)}$ is constant on the horizon. In order that the potential
be constant for the entire equivalence class $[\,\xi\ell\,]$ requires the gauge
fixing functions to be related as $\lie_{(\xi \ell)}\chi_{(\xi\ell\,)}-\Phi_{(\xi\ell\,)}
=\xi[\lie_{\ell}\chi_{(\ell\,)}-\Phi_{(\ell\,)}]$.

\vskip 0.5cm
\noindent{\em First Law}\,: The Lagrangian four-form for the electromagnetic field
is given by $8\pi L=-{\bf F}\wedge*{\bf F}$. The variation for the Lagrangian is
carried out such that ${\bf A}$ has the expected asymptotic fall-off and is gauge
fixed at $\Delta$ as in (\ref{lieema}). Here the key point is that although we
needed a gravitational surface term in the action, thanks to the electromagnetic
zeroth law, such a term is not needed here. Proceeding as in the gravitational case
here also we find a bulk and a surface term in the symplectic structure. The surface
term can be made explicit by introducing a potential for $\Phi_{(\xi\ell\,)}$:
$\lie_{(\xi\ell\,)}\varphi_{(\xi\ell\,)}\triangleq-\Phi_{(\xi\ell\,)}$. It too
suffers from an additive ambiguity which is removed by choosing
$\varphi_{(\xi\ell\,)}|_{S_-} \triangleq 0$. Then the symplectic structure is
\begin{eqnarray}\label{totalsymplectic}
 &&\Omega_{\rm em}(\delta_1,\delta_2)=
-\frac{1}{4\pi}\!\int_M\!\big[\delta_1\!*{\bf F}\wedge\delta_2{\bf A}
       -(1\leftrightarrow 2)\big]+\nn &&
 \frac{1}{4\pi}\oint_{S_\Delta}\!\big[\delta_1\!*{\bf F}\,\delta_2
 (\chi_{(\xi\ell\,)}+\varphi_{(\xi\ell\,)})-(1\leftrightarrow 2)\big]\;.
\end{eqnarray}
Again, we look for evaluating $X_t(\delta)$ from the electromagnetic part of the
symplectic structure. Making use of the field equations the bulk symplectic
structure receives contributions from the boundary alone, which is found to be
$(\lie_t\chi_{(\xi\ell\,)}-\Phi_{(\xi\ell\,)})\,\delta Q_\Delta$. To evaluate the
contribution from the surface symplectic structure care must be taken not to
identify $\delta_t$ with $\lie_t$ for the derived potential $\varphi$. It turns out
that $\delta_t\varphi_{(\xi\ell\,)}\triangleq 0$ everywhere and the nonzero
contribution is $-\delta_t\chi_{(\xi\ell\,)}$. Combining the bulk and the surface
contributions we find that
\begin{eqnarray}\label{oneformtotal}
X_t^{\rm em}(\delta)\triangleq-\Phi_{(\xi\ell\,)}\delta Q_\Delta\;.
\end{eqnarray}

In this letter we sketched how the boundary conditions defining a WIH can be relaxed for
a larger equivalence class of null normals so that still the zeroth and the first laws of
black hole mechanics emerge. The framework should be extendible to include rotations as
well. Since this formulation is blind to extremal and non-extremal global solutions it is
believed that the entropy issue can be addressed more squarely if one quantizes this WIH.
Generalizations to dynamical horizons and their settling down to such WIH would also be
addressed in the future. 

%%%%%%%%%%%%%%%%%%%%%%%%%%%%%%%%%%%%%%%%%%%%%%%%%%%%%%%%%%%%%%%%%
\begin{acknowledgments}

We thank Abhay Ashtekar for discussion and clarification of a point from the 
reference \cite{afk}. We thank Ashok Chatterjee for discussion. We especially 
thank Parthasarathi Majumdar for several stimulating discussions and continuous
encouragement during the course of the work. 

\end{acknowledgments}

%%%%%%%%%%%%%%%%%%%%%%%%%%%%%%%%%%%%%%%%%%%%%%%%%%%%%%%%%%%%%%%%%

%%%%%%%%%%%%%%%%%%%%%%%%%%%%%%%%%%%%%%%%%%%%%%%%%%%%%%%%%%%%%%%%%%

\end{document}